%% file: main.tex
\documentclass[ALICE,manyauthors]{cernphprep}
\usepackage[comma,square,numbers,sort&compress]{natbib}
\usepackage{hyperref}
\usepackage{lineno}
\usepackage{xspace}
\usepackage{comment}
\usepackage{multirow}
\usepackage{xcolor} %textcolor
\RequirePackage[normalem]{ulem} %strikethrough
\usepackage[T1]{fontenc}
\usepackage{booktabs}

\begin{document}
\input{commands.tex}

%%%%%%%%%%%%%%%  Title page %%%%%%%%%%%%%%%%%%%%%%%%
\begin{titlepage}
% the dates below correspond to CERN approval
% please don't touch: EB chairs will take care
\PHyear{2021}       % required, will be obtained from CERN
\PHnumber{260}      % required, will be obtained from CERN
\PHdate{15 December}  % required, will be obtained from CERN
%%%%%%%%%%%%%%%%%%%%%%%%%%%%%%%%%%%%%%%%%%%%%%%%%%%%

%%% Put your own title + short title here:
\title{Constraining hadronization mechanisms with \LcD production ratios in Pb--Pb collisions at $\sqrt{s_{\rm NN}} = 5.02$~TeV}
\ShortTitle{\Lc production in Pb--Pb collisions at $\sqrt{s_{\rm NN}} = 5.02$~TeV}   % appears on left page headers

%%% Do not change the next lines
\Collaboration{ALICE Collaboration\thanks{See Appendix~\ref{app:collab} for the list of collaboration members}}
\ShortAuthor{ALICE Collaboration} % appears on right page headers, do not change

\begin{abstract}
\input{0_Abstract}
\end{abstract}
\end{titlepage}

\setcounter{page}{2} %please do not remove this line

%%%%%%%%%%%%%%%%%%%%%%%%%%%%%%%%
% begin main text
%%%%%%%%%%%%%%%%%%%%%%%%%%%%%%%%
\input{1_Introduction}

\input{2_Apparatus_Data}

\input{3_Analysis}

\input{4_Results}

\input{5_Conclusion}
%%%%%%%%%%%%%%%%%%%%%%%%%%%%%%%%
% end main text 
%%%%%%%%%%%%%%%%%%%%%%%%%%%%%%%%

%%%%% acknowledgements - handled by EB chairs 
\newenvironment{acknowledgement}{\relax}{\relax}
\begin{acknowledgement}
\section*{Acknowledgements}
% add specific acknowledgements here 
% ...but please don't remove the line below: funding agencies
% will be acknowledged with a custom tex file handled by EB chairs after Collab Round 2
\input{fa_2021-11-30.tex}
\end{acknowledgement}

%%%%%%%% Bibliography 
\bibliographystyle{utphys}   % Remember we use title in the biblio
\bibliography{bibliography}
%\input {bibliography.tex}  

%%%%%%%%%%%%%%%%%%%%%%%%%%%%%%%%
% Appendices: yours (if any) + authorlist
%%%%%%%%%%%%%%%%%%%%%%%%%%%%%%%%
\newpage
\appendix

\input{6_SupplementalFigures}
\clearpage

%%%%% Authorlist - please do not touch: handled by EB chairs 
\section{The ALICE Collaboration}
\label{app:collab}
\input{2021-11-30-Alice_Authorlist_2021-11-30.tex}

\end{document}

%% file: commands.tex
%%%%%%%%%%%%%%%%%%%%%%%%%%%%%%%%%%%%%%%%%%%%%%%%%%
% These are some new commands that may be useful 
% for paper writing in general. If other newcommands
% are needed for your specific paper, please feel 
% free to add here. 
%
% The currently available commands are organized in: 
% 1) Systems
% 2) Quantities
% 3) Energies and units
% 4) Detectors
% 5) particle species 
%%%%%%%%%%%%%%%%%%%%%%%%%%%%%%%%%%%%%%%%%%%%%%%%%%

\newcommand{\red}[1]{\textcolor{red}{#1}}
\newcommand{\blue}[1]{\textcolor{blue}{#1}}
\newcommand{\orange}[1]{\textcolor{orange}{#1}}

% 1) SYSTEMS 
\newcommand{\pp}           {pp\xspace}
\newcommand{\pplong}       {proton-proton\xspace}
\newcommand{\ppbar}        {\mbox{$\mathrm {p\overline{p}}$}\xspace}
\newcommand{\ee}           {\mbox{$\mathrm {e^{+}e^{-}}$}\xspace}
\newcommand{\ep}           {\mbox{$\mathrm {e^{-}p}$}\xspace}
\newcommand{\XeXe}         {\mbox{Xe--Xe}\xspace}
\newcommand{\PbPb}         {\mbox{Pb--Pb}\xspace}
\newcommand{\pA}           {\mbox{pA}\xspace}
\newcommand{\pPb}          {\mbox{p--Pb}\xspace}
\newcommand{\AuAu}         {\mbox{Au--Au}\xspace}
\newcommand{\dAu}          {\mbox{d--Au}\xspace}
\let\eepm=\ee
\newcommand{\hi}           {heavy-ion\xspace}
\newcommand{\Hi}           {Heavy-ion\xspace}

% 2) QUANTITIES 
\newcommand{\s}            {\ensuremath{\sqrt{s}}\xspace}
\newcommand{\snn}          {\ensuremath{\sqrt{s_{\mathrm{NN}}}}\xspace}
\newcommand{\pt}           {\ensuremath{p_{\rm T}}\xspace}
\newcommand{\kt}           {\ensuremath{k_{\rm T}}\xspace}
\newcommand{\meanpt}       {$\langle p_{\mathrm{T}}\rangle$\xspace}
\newcommand{\ycms}         {\ensuremath{y_{\rm CMS}}\xspace}
\newcommand{\ylab}         {\ensuremath{y_{\rm lab}}\xspace}
\newcommand{\etarange}[1]  {\mbox{$\left | \eta \right |~<~#1$}\xspace}
\newcommand{\yrange}[1]    {\mbox{$\left | y \right |~<~#1$}\xspace}
\newcommand{\dndy}         {\ensuremath{\mathrm{d}N_\mathrm{ch}/\mathrm{d}y}\xspace}
\newcommand{\dndeta}       {\ensuremath{\mathrm{d}N_\mathrm{ch}/\mathrm{d}\eta}\xspace}
\newcommand{\avdndeta}     {\ensuremath{\langle\dndeta\rangle}\xspace}
\newcommand{\dNdy}         {\ensuremath{\mathrm{d}N_\mathrm{ch}/\mathrm{d}y}\xspace}
\newcommand{\Npart}        {\ensuremath{N_\mathrm{part}}\xspace}
\newcommand{\Ncoll}        {\ensuremath{N_\mathrm{coll}}\xspace}
\newcommand{\dEdx}         {\ensuremath{\textrm{d}E/\textrm{d}x}\xspace}
\newcommand{\DeltaM}       {\ensuremath{\Delta M}\xspace}
\let\dNchdeta=\avdndeta
\newcommand{\ntrkl}        {\ensuremath{N_\mathrm{trkl}}\xspace}
\newcommand{\vzeromperc}   {\ensuremath{p_\mathrm{V0M}}\xspace}
\newcommand{\zvtx}         {\ensuremath{z_\mathrm{vtx}}\xspace}
\newcommand{\raa}          {\ensuremath{R_\mathrm{AA}}\xspace}
\newcommand{\taa}          {\ensuremath{\langle T_{\rm AA}\rangle}\xspace}
\newcommand{\RPbPb}         {\ensuremath{R_\mathrm{PbPb}}\xspace}
\newcommand{\RpPb}         {\ensuremath{R_\mathrm{pPb}}\xspace}
\let\RAA=\raa
\let\Raa=\raa
\let\TAA=\taa
\let\Taa=\taa

% 3) ENERGIES, UNITS
\newcommand{\sqrts}        {\ensuremath{\sqrt{s}}\xspace}
\newcommand{\sqrtsNN}      {\ensuremath{\sqrt{s_{\mathrm{NN}}}}\xspace}
\newcommand{\nineH}        {$\sqrt{s}=0.9$~Te\kern-.1emV\xspace}
\newcommand{\seven}        {$\sqrt{s}=7$~Te\kern-.1emV\xspace}
\newcommand{\twoH}         {$\sqrt{s}=0.2$~Te\kern-.1emV\xspace}
\newcommand{\twosevensix}  {$\sqrt{s}=2.76$~Te\kern-.1emV\xspace}
\newcommand{\five}         {$\sqrt{s}=5.02$~Te\kern-.1emV\xspace}
\newcommand{\thirteen}     {$\sqrt{s}=13$~Te\kern-.1emV\xspace}
\newcommand{\twoHnn}       {$\sqrt{s_{\mathrm{NN}}}=200$~Ge\kern-.1emV\xspace}
\newcommand{\twosevensixnn}{$\sqrt{s_{\mathrm{NN}}}=2.76$~Te\kern-.1emV\xspace}
\newcommand{\fivenn}       {$\sqrt{s_{\mathrm{NN}}}=5.02$~Te\kern-.1emV\xspace}
\newcommand{\LT}           {L{\'e}vy-Tsallis\xspace}
\newcommand{\GeVc}         {Ge\kern-.1emV$/c$\xspace}
\newcommand{\MeVc}         {Me\kern-.1emV$/c$\xspace}
\newcommand{\TeV}          {Te\kern-.1emV\xspace}
\newcommand{\GeV}          {Ge\kern-.1emV\xspace}
\newcommand{\MeV}          {Me\kern-.1emV\xspace}
\newcommand{\GeVcc}        {Ge\kern-.1emV$/c^2$\xspace}
\newcommand{\MeVcc}        {Me\kern-.1emV$/c^2$\xspace}
\newcommand{\lumi}         {\ensuremath{\mathcal{L}}\xspace}
\newcommand{\degree}       {\ensuremath{^{\rm o}}\xspace}
\newcommand{\inversemub}   {\ensuremath{\mu \rm b^{\rm -1}}\xspace}
\newcommand{\inversemubnew}{µb\ensuremath{^{\rm -1}}\xspace}

\let\gevc=\GeVc
\let\mevc=\MeVc
\let\tev=\TeV
\let\gev=\GeV
\let\mev=\MeV
\let\gevcc=\GeVcc
\let\mevcc=\MeVcc
\let\GeVmom=\GeVc
\let\MeVmom=\MeVc
\let\GeVmass=\GeVcc
\let\MeVmass=\MeVcc

% 4) DETECTORS 
\newcommand{\ITS}          {\rm{ITS}\xspace}
\newcommand{\TOF}          {\rm{TOF}\xspace}
\newcommand{\ZDC}          {\rm{ZDC}\xspace}
\newcommand{\ZDCs}         {\rm{ZDCs}\xspace}
\newcommand{\ZNA}          {\rm{ZNA}\xspace}
\newcommand{\ZNC}          {\rm{ZNC}\xspace}
\newcommand{\SPD}          {\rm{SPD}\xspace}
\newcommand{\SDD}          {\rm{SDD}\xspace}
\newcommand{\SSD}          {\rm{SSD}\xspace}
\newcommand{\TPC}          {\rm{TPC}\xspace}
\newcommand{\TRD}          {\rm{TRD}\xspace}
\newcommand{\VZERO}        {\rm{V0}\xspace}
\newcommand{\VZEROA}       {\rm{V0A}\xspace}
\newcommand{\VZEROC}       {\rm{V0C}\xspace}

% 4) PARTICLE SPECIES 
\newcommand{\pip}          {\ensuremath{\pi^{+}}\xspace}
\newcommand{\pim}          {\ensuremath{\pi^{-}}\xspace}
\newcommand{\kap}          {\ensuremath{\rm{K}^{+}}\xspace}
\newcommand{\kam}          {\ensuremath{\rm{K}^{-}}\xspace}
\newcommand{\pbar}         {\ensuremath{\rm\overline{p}}\xspace}
\newcommand{\kzero}        {\ensuremath{{\rm K}^{0}_{\rm{S}}}\xspace}
\newcommand{\lmb}          {\ensuremath{\Lambda}\xspace}
\newcommand{\almb}         {\ensuremath{\overline{\Lambda}}\xspace}
\newcommand{\Om}           {\ensuremath{\Omega^-}\xspace}
\newcommand{\Mo}           {\ensuremath{\overline{\Omega}^+}\xspace}
\newcommand{\X}            {\ensuremath{\Xi^-}\xspace}
\newcommand{\Ix}           {\ensuremath{\overline{\Xi}^+}\xspace}
\newcommand{\Xis}          {\ensuremath{\Xi^{\pm}}\xspace}
\newcommand{\Oms}          {\ensuremath{\Omega^{\pm}}\xspace}
\newcommand{\Vdecay} 	   {\ensuremath{\rm V^{0}}\xspace}
\newcommand{\Kzeros}       {\ensuremath{\rm K^{0}_{\rm S}}\xspace}
\let\Kzs=\Kzeros

% 5) CUSTOM
\newcommand{\Dzero}        {\ensuremath{\rm D^{0}}\xspace}
\newcommand{\Dplus}        {\ensuremath{\rm D^{+}}\xspace}
\newcommand{\Dstar}        {\ensuremath{\rm D^{*+}}\xspace}
\newcommand{\Ds}           {\ensuremath{\rm D^{+}_{\rm s}}\xspace}
\newcommand{\Lambdac}      {\ensuremath{\rm \Lambda_{\rm c}^{+}}\xspace}
\newcommand{\Bzero}        {\ensuremath{\rm B^{0}}\xspace}
\newcommand{\Bplus}        {\ensuremath{\rm B^{+}}\xspace}
\newcommand{\Bs}           {\ensuremath{\rm B^{+}_{\rm s}}\xspace}
\newcommand{\Lambdab}      {\ensuremath{\rm \Lambda_{\rm b}^{0}}\xspace}
\let\Lc=\Lambdac
\let\Lb=\Lambdab

\newcommand{\DtoKpi}          {\ensuremath{\rm D^{0} \to K^{-}\pi^{+}}\xspace}
\newcommand{\Dstophip}        {\ensuremath{\rm D_{\rm s}^{+} \to \phi \pi^{+} \to K^{+}K^{-}\pi^{+}}\xspace}
\newcommand{\LctopKpi}        {\ensuremath{\rm \Lambda_{\rm c}^{+} \to pK^{-}\pi^{+}}\xspace}
\newcommand{\LctopKzeros}     {\ensuremath{\rm \Lambda_{\rm c}^{+} \to pK^{0}_{\rm S}}\xspace}
\newcommand{\Lcpm}       {\ensuremath{\rm \Lambda_{\rm c}^{\pm}}\xspace}
\newcommand{\LctopKzerosfull} {\ensuremath{\rm \Lambda_{\rm c}^{+} \to pK^{0}_{\rm S} \to p\pi^{+}\pi^{-}}\xspace}
\newcommand{\Kzerostopipi} {\ensuremath{\rm K^{0}_{\rm S} \to \pi^{+}\pi^{-}}\xspace}
\newcommand{\pKpi}            {\ensuremath{\rm pK^{-}\pi^{+}}\xspace}
\newcommand{\pKzeros}         {\ensuremath{\rm pK^{0}_{\rm S}}\xspace}
\let\LambdactopKpi=\LctopKpi
\let\LambdactopKzeros=\LctopKzeros
\let\LambdactopKzerosfull=\LctopKzerosfull

\newcommand{\DsDzero}         {\ensuremath{\Ds/\Dzero}\xspace}
\newcommand{\LcDzero}         {\ensuremath{\Lc/\Dzero}\xspace}
\let\LcD=\LcDzero
\newcommand{\inel}            {\ensuremath{\rm INEL_{>0}}\xspace}
\let\INEL=\inel

\newcommand{\Hf}            {Heavy-flavour\xspace}
\newcommand{\hf}            {heavy-flavour\xspace}

\newcommand{\alphas}        {\ensuremath{\alpha_\mathrm{S}\xspace}}

% 6) DOCUMENT, STRUCTURE, HELPERS
\newcommand{\secletter}[1]    {\vspace{0.3cm}\textit{#1}}

\newcommand{\RAAprompt}    {\ensuremath{R^{\rm prompt}_\mathrm{AA}}\xspace}
\newcommand{\RAAnonprompt}    {\ensuremath{R^{\rm non\text{-}prompt}_\mathrm{AA}}\xspace}

%% file: 0_Abstract.tex
The production of prompt \Lc baryons at midrapidity ($|y|<0.5$) was measured in central (0--10\%) and mid-central (30--50\%) \PbPb collisions at the center-of-mass energy per nucleon--nucleon pair \fivenn with the ALICE detector. The results are more precise, more differential in centrality, and reach much lower transverse momentum ($\pt=1$~\gevc) with respect to previous measurements performed by the ALICE, STAR, and CMS Collaborations in nucleus--nucleus collisions, allowing for an extrapolation down to $\pt=0$. The \pt-differential \LcD ratio is enhanced with respect to the \pp measurement for $4<\pt<8$~\gevc by 3.7 standard deviations ($\sigma$), while the \pt-integrated ratios are compatible within 1$\sigma$. The observed trend is similar to that observed in the strange sector for the $\Lambda/\kzero$ ratio. Model calculations including coalescence or statistical hadronization for charm-hadron formation are compared with the data.

%% file: 1_Introduction.tex
\section{Introduction}

Heavy-ion collisions at LHC energies produce a phase of strongly-interacting matter, known as the quark--gluon plasma (QGP), in which quarks and gluons are deconfined~\cite{Busza:2018rrf}. The existing measurements indicate that the QGP behaves as a strongly-coupled low-viscosity liquid-like system~\cite{Muller:2012zq}. Heavy quarks, produced at the start of the collision, experience the full evolution of the system and constitute a unique probe of the QGP properties~\cite{Andronic:2015wma}. The hadronization of heavy quarks into open heavy-flavor hadrons is expected to be influenced by the presence of a deconfined medium. Theoretical calculations that include modified hadronization via quark coalescence or via a resonance recombination approach~\cite{Lee:2007wr,Oh:2009zj,Das:2016llg,Plumari:2017ntm,He:2019vgs,Beraudo:2022dpz} predict a significant enhancement of the \LcD yield ratio in heavy-ion collisions compared to the expected ratio in \pp collisions. In addition, the collective radial expansion of the system determines a flow-velocity profile common to all thermalized particles, that could increase the \LcD ratio at intermediate transverse momentum, i.e.~$2 \lesssim \pt \lesssim 8$~\gevc~\cite{Plumari:2017ntm,Andronic:2021erx,He:2019vgs}. The study of such a potential enhancement requires a good understanding of \Lc production in smaller collision systems, which showed surprising features at LHC energies. The production of \Lambdac baryons was measured at the LHC in \pp collisions by the ALICE Collaboration at $\sqrts = 5.02$, 7, and 13~TeV~\cite{Acharya:2020lrg,Acharya:2020uqi,Acharya:2017kfy,Acharya:2021vpo}, by the CMS Collaboration at 5.02~TeV~\cite{Sirunyan:2019fnc}, and by the LHCb Collaboration at 7~TeV~\cite{LHCb:2013xam}. At midrapidity, the ALICE and CMS results show a significant enhancement in the \LcDzero yield ratio (up to a factor 2--5 for $\pt<8$~\gevc) compared to \ee and \ep measurements~\cite{Albrecht:1988an, Avery:1990bc, Albrecht:1991ss, Chekanov:2005mm, Abramowicz:2013eja, Abramowicz:2010aa} and QCD-inspired theoretical predictions~\cite{Sjostrand:2014zea, Bahr:2008pv, Frixione:2007nw, Kniehl:2020szu} where charm fragmentation is tuned on \ee and \ep measurements~\cite{Gladilin:2014tba, Zyla:2020zbs}. Models that introduce new color-reconnection topologies in string fragmentation~\cite{Christiansen:2015yqa} or hadron production via coalescence~\cite{Minissale:2020bif} as well as models that are based on statistical hadronization including feed-down from unobserved charm-baryon states~\cite{He:2019tik} are able to describe the \LcD ratio at midrapidity. The values of the \LcD ratio measured at forward rapidity by the LHCb Collaboration are smaller than those at midrapidity, indicating a non-negligible rapidity dependence.

A recent measurement performed by ALICE in intervals of charged-particle multiplicity \dndeta in \pp collisions at \thirteen~\cite{ALICE:2021npz} showed that in a hadronic collision, even at relatively small multiplicities, charm-quark hadronization proceeds differently than in \ee collisions. The \pt-dependence of the \LcD ratio evolves with multiplicity and the maximum of the ratio increases for the higher multiplicity intervals, while the \pt-integrated \LcD ratios do not show a significant dependence on multiplicity up to $\avdndeta \approx 40$. Whether the \pt-differential \LcD ratio keeps evolving with multiplicity up to the typical multiplicities of \PbPb collisions, and whether an overall \pt-integrated enhancement of \Lc production relative to the \Dzero one is present at higher multiplicities, as proposed by coalescence models including light diquark states~\cite{Lee:2007wr,Oh:2009zj,Beraudo:2022dpz}, are open questions and fundamental to the understanding of charm-quark hadronization.

The \Lambdac production in nucleus--nucleus collisions was measured for the first time at the LHC by ALICE in \PbPb collisions at \fivenn in the 0--80\% centrality interval for $6<\pt<12$~\gevc~\cite{Acharya:2018ckj}. The \LcD ratio was found to be close to unity, larger than the corresponding ratio measured in \pp collisions, and well described by calculations including hadronization via coalescence mechanisms~\cite{Plumari:2017ntm,He:2019vgs}. The \LcD ratio measured in the interval $3<\pt<6$~\gevc by the STAR Collaboration in \AuAu collisions at \twoHnn~\cite{Adam:2019hpq} shows an increasing trend towards more central collisions and is also described by model calculations including hadronization via coalescence~\cite{Oh:2009zj,Cho:2019lxb,Plumari:2017ntm,Zhao:2018jlw,He:2019vgs,Cao:2019iqs}. Considering together the values calculated by STAR in 10--80\% \AuAu collisions and by ALICE in \pp and \pPb collisions~\cite{Acharya:2020lrg,ALICE:2021npz} a possible increase of the \pt-integrated \LcD ratio at high multiplicity is neither excluded nor confirmed. The CMS measurement in \PbPb collisions at \fivenn~\cite{Sirunyan:2019fnc}, performed in the interval $10<\pt<20$~\gevc, is consistent with the \pp result within uncertainties as well as with predictions considering only string fragmentation~\cite{Christiansen:2015yqa}, suggesting that coalescence has no significant effect in this \pt range. The production of \Lc baryons was also measured in \pPb collisions at \fivenn by the ALICE and LHCb Collaborations~\cite{Acharya:2020lrg,Acharya:2020uqi,LHCb:2018weo}. The current measurements do, however, not allow to draw conclusions on the role of different cold-nuclear matter effects and the possible presence of hot-medium effects. Recently, LHCb also measured the \LcD ratio in peripheral (65--90\%) \PbPb collisions at \fivenn, which was observed to be consistent with the LHCb ratio in \pPb collisions~\cite{LHCb:2022ddg}.

In this letter, the measurement of the \pt-differential production yields of prompt \Lc baryons in central (0--10\%) and mid-central (30--50\%) collisions using the 2018 \PbPb at \fivenn are reported down to $\pt = 1$~\gevc. The results are more precise and more differential in \pt and centrality with respect to previous measurements~\cite{Acharya:2018ckj,Adam:2019hpq,Sirunyan:2019fnc}. The \LcD yield ratios and the nuclear modification factor \Raa, which is defined as the ratio of the production yield in \PbPb collisions and the cross section in \pp collisions scaled by the average nuclear overlap function \TAA (proportional to the number of nucleon--nucleon collisions), are reported as function of \pt and compared with theoretical predictions. The \pt-integrated \Lc production yield and \LcD ratio, extrapolated to $\pt=0$, are also presented for the first time in \PbPb collisions.

%% file: 2_Apparatus_Data.tex
\section{Experimental apparatus and data sample}

The ALICE apparatus is described in detail in~\cite{Aamodt:2008zz,Abelev:2014ffa}. The data were collected using triggers based on the signal in the \VZERO detectors~\cite{Abbas:2013taa}. A minimum bias trigger, which required coincident signals in both detecting components of the \VZERO detector along the beam axis on opposite sides of the interaction point, was exploited. In addition, and differently with respect to the previous \PbPb data taking period at the same \snn, two new trigger selections were introduced to enrich the sample of central and mid-central collisions via an online event selection based on the \VZERO-signal amplitude. Events were further selected offline using timing information from the \VZERO detectors and the neutron Zero Degree Calorimeters~\cite{Arnaldi:1999zz} to reject events due to the interaction of one of the beams with residual gas in the vacuum tube. Furthermore, only events with a primary vertex reconstructed within $\pm10$~cm from the center of the detector along the beam axis were considered in the analysis. Collisions were classified into centrality intervals, defined in terms of percentiles of the hadronic \PbPb cross section, using the \VZERO-signal amplitudes~\cite{Acharya:2018hre}. The number of events in the centrality classes 0--10\% and 30--50\% considered for this analysis is about $100 \times 10^6$ and $85 \times 10^{6}$, respectively, corresponding to a luminosity of $(130.5\pm0.5)$~{\inversemub} and $(55.5\pm0.2)$~{\inversemub}~\cite{ALICE:2018tvk}. 

The Monte Carlo (MC) simulations utilized in this analysis were obtained using the HIJING~1.36 event generator~\cite{Wang:1991hta} to simulate \PbPb collisions at \fivenn. In each simulated event, \Lc signals were added by injecting $\rm c\overline{c}$ or $\rm b\overline{b}$ pairs generated with the PYTHIA~8.243 event generator~\cite{Sjostrand:2014zea} with the Monash tune~\cite{Skands:2014pea}. The \Lc baryons were forced to decay into the hadronic decay channel of interest, \LctopKzeros followed by \Kzerostopipi, using PYTHIA. All generated particles were transported through the ALICE detector using the GEANT3 package~\cite{Brun:1994aa}. The conditions of all the ALICE detectors in terms of active channels, gain, noise level, and alignment, and their evolution with time during the data taking, were taken into account in the simulations.

%% file: 3_Analysis.tex
\section{Data analysis}

The \Lambdac baryon and its charge conjugate were reconstructed by exploiting the topology of the hadronic decay channel \LctopKzeros (branching ratio ${\rm BR} = 1.59 \pm 0.08 \% $), followed by the subsequent decay \Kzerostopipi (${\rm BR} = 69.20 \pm 0.05 \% $)~\cite{Zyla:2020zbs}. Charged-particle tracks used to define the \Lc candidates are reconstructed using the Inner Tracking System (\ITS)~\cite{Aamodt:2010aa} and the Time Projection Chamber (\TPC)~\cite{Alme:2010ke}, located in a solenoid magnet that provides a $0.5$~T field parallel to the beam direction. The \LctopKzeros candidates combine a proton-candidate track with a \Kzs-meson candidate, reconstructed in the \Kzerostopipi decay channel. Only proton (pion) tracks with $|\eta|<0.8$, $\pt>0.4\ (0.1)$~\gevc, at least 70 out of 159 associated crossed TPC pad rows, a ratio of crossed rows to findable clusters in the TPC larger than 0.8, at least 50 clusters in the TPC available for particle identification (PID), and a $\chi^2/{\rm ndf} < 1.25$ in the TPC (where $\rm ndf$ is the number of degrees of freedom involved in the track fit procedure) were considered for the analysis. Moreover, a minimum number of two hits (out of six) in the ITS, with at least one in the inner two layers, were required for the proton track. The selection of tracks with $|\eta|<0.8$ limits the \Lambdac acceptance in rapidity. For this reason a fiducial acceptance selection was applied on the rapidity of the \Lambdac candidates, $|y_\mathrm{lab}| < y_\mathrm{fid}(\pt)$, where $y_\mathrm{fid}$ increases from 0.6 to 0.8 in $1 < \pt < 5$~\gevc, and $y_\mathrm{fid} = 0.8$ for $\pt > 5$~\gevc.

Unlike the previous analysis based on linear selections~\cite{Acharya:2018ckj}, the \Lambdac-candidate selection was performed using multivariate techniques based on the Boosted Decision Tree (BDT) algorithm provided by the XGBoost package~\cite{Chen:2016:XST:2939672.2939785}. Before the training, loose kinematic and topological selections were applied to the \kzero-meson candidate together with the particle identification of the proton-candidate track. The PID was performed using the specific ionization energy loss \dEdx in the TPC gas and the time of flight from the interaction point to the Time-Of-Flight (TOF) detector~\cite{Akindinov:2013tea,ALICE:2016ovj}. The BDT training was performed considering as signal candidates prompt (not coming from beauty-hadron decays) \Lambdac decays from MC simulations. Background candidates were taken from the sidebands of the invariant mass distribution in data (defined to be outside a 80~\MeVcc window around the \Lc mass value reported by the PDG~\cite{Zyla:2020zbs}).

The variables that were most important in the training were the PID-related variables of the proton-candidate track, the displacement of the proton-candidate track from the primary vertex, the distance between the \kzero-meson decay vertex and the primary vertex, and the cosine of the pointing angle between the \kzero-meson candidate line of flight and its reconstructed momentum vector. Independent BDTs were trained for the different \pt and centrality intervals.

The selection on the BDT output was tuned in each \pt interval to maximize the expected statistical significance, which is estimated using i) the expected signal obtained from FONLL calculations~\cite{Cacciari:1998it,Cacciari:2012ny} scaled by the corresponding \TAA~\cite{ALICE:2018tvk} and multiplied by the BDT selection efficiency and ii) the expected background estimated from an invariant mass sideband fit using a fraction of the data.

After applying selections on the BDT output, the yields of \Lc baryons were extracted in each \pt interval via binned maximum-likelihood fits to the candidate invariant mass distributions. The fitting function consisted of a Gaussian term to estimate the signal and a second-, third-, or fourth-order polynomial function (depending on the \pt interval) to estimate the background. The default background fitting function was chosen after dedicated studies to obtain a good description of the invariant mass distribution in the sidebands. The other functions were considered for evaluating the systematic uncertainty.

The raw-yield extraction is challenging, especially at low \pt with signal-to-background ratios below one per mille and relative statistical uncertainties on the extracted raw yield varying between 15--35\%, as presented in Appendix~\ref{app:Mass}. Given the critical signal extraction due to the low signal-to-background ratios, the width of the Gaussian term for the signal was fixed to the value obtained from simulations. It was verified that the widths from the simulation were consistent within uncertainties to those extracted from fits to data without constraints on the width of the Gaussian (with a relative uncertainty of 1--2\% in simulation and 20--30\% in data). In addition, the stability of the signal extraction was further verified by i) fitting purely background candidates from simulations and ii) by repeating the fit after subtracting a background component estimated with an event-mixing technique. For the latter, the events were grouped in pools based on the primary-vertex position along $z$ and the estimated centrality. For the first study, none of the invariant mass fits allowed to extract a signal in the \Lc invariant mass region. For the second study, fits to the background-subtracted invariant mass distributions resulted in compatible \Lc raw yields to the ones extracted from the default fits.

The corrected yields of prompt \Lc baryons were obtained in each centrality interval as
\begin{equation}
\left. \frac{{\rm d}N^{\Lc}}{{\rm d}\pt} \right|_{|y| < 0.5} = \frac{f_{\rm{prompt}} \times \frac{1}{2} \left.N^{\rm
    \Lcpm}_{\rm raw} \right|_{|y| < y_{\rm fid}}}{\Delta\pt \times c_{\Delta y} \times (A \times
   \epsilon)_{\rm prompt} \times {\rm BR} \times N_{\rm ev}}.
\end{equation}
The raw yield values $N^{\Lcpm}_{\rm raw}$, extracted in a given \pt interval of width $\Delta\pt$, were divided by a factor two and multiplied by the prompt fraction $f_{\rm prompt}$ to obtain the charge-averaged yields of prompt \Lc. Furthermore, they were divided by $c_{\Delta y} \times (A \times \epsilon)$, enclosing the rapidity coverage and the acceptance-times-efficiency, by the $\rm BR$ of the decay channel, and by the number of analyzed events $N_{\rm ev}$.

The $(A \times \epsilon)$ correction was determined from MC simulations, using samples not employed in the BDT training. The generated \pt spectrum used to calculate the efficiencies was reweighted to reproduce the shape obtained from the \Dzero measurement~\cite{ALICE:2021rxa} multiplied by \LcD calculations from the TAMU model~\cite{He:2019vgs} in 0--10\% and 30--50\% \PbPb collisions at \fivenn. The $(A \times \epsilon)$ increases from 1\% (3\%) at low \pt to about 12\% (16\%) at high \pt for central (mid-central) collisions. The correction factor for the rapidity acceptance, $c_{\Delta y}$, was computed as the ratio between the generated \Lc-baryon yield in $\Delta y = 2 y_{\rm fid}(\pt)$ and that in $|y|<0.5$ using the reweighted \pt shape and the rapidity distribution from PYTHIA~8 simulations~\cite{Sjostrand:2014zea}. It was verified in~\cite{ALICE:2021rxa} that for ${\rm D}$ mesons the calculation of $c_{\Delta y}$ is only weakly sensitive to the rapidity distribution used for its calculation.

The $f_{\rm prompt}$ fraction of the reconstructed signal was estimated using a similar strategy as described in~\cite{Acharya:2018ckj}. In particular, the beauty-hadron production cross section was estimated with FONLL calculations~\cite{Cacciari:1998it,Cacciari:2012ny}, the fraction of beauty quarks that fragment into \Lb was estimated from the $\Lb/(\Bzero+\Bplus)$ ratio measured by LHCb in \pp collisions at \thirteen~\cite{Aaij:2019pqz} following the same strategy as used in~\cite{Acharya:2020lrg}, and the kinematics of the decay of beauty hadrons $H_{\rm b} \rightarrow \Lc + X$ simulated with PYTHIA~8~\cite{Sjostrand:2014zea}. The branching ratios were taken as implemented in PYTHIA~8.243, corresponding to approximately 82\% for \Lb baryons and 2\% for either \Bzero, \Bplus, and \Bs mesons. In addition, the $f_{\rm prompt}$ fraction is modified to account for the nuclear modification factor of \Lc baryons from beauty-hadron decays. The central correction is chosen such that $\RAAnonprompt = 2 \times \RAAprompt$ as predicted by the ``Catania'' theoretical calculation~\cite{Das:2016llg}. The resulting $f_{\rm prompt}$ fraction was found to be about 0.97 at low \pt and about 0.81 at high \pt.

The systematic uncertainties of the \Lc corrected yields include contributions from i) the extraction of the raw yield, ii) the tracking efficiency, iii) the \Lc selection efficiency, iv) the MC generated \pt spectra, v) the statistical uncertainty of the efficiency, and vi) the subtraction of feed-down \Lc baryons from b-hadron decays. The estimated values of these systematic uncertainties are summarized for representative \pt intervals in Table~\ref{tab:syst}. In addition, a global systematic uncertainty due to the centrality interval definition (2\% for mid-central, negligible for central)~\cite{ALICE:2021rxa} and the branching ratio (5.5\%)~\cite{Zyla:2020zbs} was assigned. For the \RAA observable, the uncertainty of the \pp cross section normalization uncertainty (2.1\%)~\cite{Acharya:2020lrg} and of the average nuclear overlap function (0.7\% for central, 1.6\% for mid-central)~\cite{ALICE:2018tvk} are included in the global normalization uncertainty.

\begin{table}[tb]
    \caption{Relative systematic uncertainties of the prompt \Lc-baryon corrected yield in \PbPb collisions for central and mid-central events in representative \pt intervals.}
    \centering
    \renewcommand*{\arraystretch}{1.4}
    \begin{tabular}[t]{l|>{\centering}p{0.05\linewidth}>{\centering}p{0.05\linewidth}|>{\centering}p{0.05\linewidth}>{\centering}p{0.05\linewidth}}
        \toprule
        Centrality interval & \multicolumn{2}{c|}{0--10\%} & \multicolumn{2}{c}{30--50\%} \\
        \pt~(\gevc) & 4--6 & \multicolumn{1}{c|}{12--24} & 1--2 & \multicolumn{1}{c}{6--8} \\
        \midrule
        Yield extraction & 11\% & \multicolumn{1}{c|}{17\%} & 14\% & \multicolumn{1}{c}{12\%} \\
        Tracking efficiency & 10\% & \multicolumn{1}{c|}{9\%} & 12\% & \multicolumn{1}{c}{8\%} \\
        Selection efficiency & 8\% & \multicolumn{1}{c|}{8\%} & 7\% & \multicolumn{1}{c}{7\%} \\
        Prompt fraction & $^{+8}_{-6}$\% & \multicolumn{1}{c|}{$^{+13}_{-13}$\%} & $^{+4}_{-3}$\% & \multicolumn{1}{c}{$^{+12}_{-8}$\%} \\
        MC \pt shape & 2\%   & \multicolumn{1}{c|}{negl.} & 2\% & \multicolumn{1}{c}{1\%} \\
        Centrality limits & \multicolumn{2}{c|}{$<0.1$\%} & \multicolumn{2}{c}{2\%} \\
        \midrule
        Branching ratio & \multicolumn{4}{c}{$5.5$\%} \\
        \midrule
        Total syst. unc. & $^{+20}_{-19}$\% & \multicolumn{1}{c|}{$^{+25}_{-25}$\%} & $^{+21}_{-21}$\% & \multicolumn{1}{c}{$^{+21}_{-19}$\%} \\
        \bottomrule
    \end{tabular}
    \label{tab:syst}
\end{table}

The systematic uncertainty of the raw-yield extraction was estimated by repeating the invariant mass fits varying the lower and upper limits of the fit range, the functional form of the background fit function, and considering the Gaussian width (mean) as a free (fixed) parameter in the fit. In order to test the sensitivity to the line shape of the signal, a bin-counting method was used, in which the signal yield was obtained by integrating the invariant-mass distribution after subtracting the background estimated from the sideband fit, as well as by studying the signal shape in the MC simulations using multiple stacked Gaussian functions rather than a single one.
The procedure to estimate the systematic uncertainty of the track-reconstruction efficiency includes variations of the track-quality selection criteria for all decay tracks and studies on the probability to match TPC tracks to the ITS clusters in data and simulation for the proton-candidate track. The latter comparison was performed after weighting the relative abundances of primary and secondary particles in simulation to those in data~\cite{Acharya:2018ckj}. The systematic uncertainty of the \Lc selection efficiency was estimated by repeating the analysis with different selections on the BDT output, resulting in up to 50\% lower and 20--50\% higher efficiency values. Possible systematic effects due to the loose PID selection, applied prior to the BDT one, were investigated by comparing the PID-selection efficiencies in data and in simulations and found to be negligible. Both the tracking- and PID-efficiency studies were performed using pure samples of pions (from \kzero decays) and protons (from $\Lambda$ decays).
An additional contribution derives from the \pt spectra of \Lc generated in the simulation, which was estimated by using the \LcD predictions of the Catania model~\cite{Plumari:2017ntm} and the SHMc~\cite{Andronic:2021erx} instead of the TAMU prediction~\cite{He:2019vgs} in the \pt-shape reweighting procedure, as well as by an iterative method using a parametrization of the measured \pt-differential production yields. 
Finally, the systematic uncertainty of the feed-down subtraction was estimated by varying the FONLL parameters as prescribed in~\cite{Cacciari:2012ny} and the function describing the \Lb fragmentation fraction within the quoted experimental uncertainty as reported in~\cite{Acharya:2020lrg}, as well as by varying the hypothesis on \RAAnonprompt. For the latter, an interval $1/3 < \RAAnonprompt/\RAAprompt < 3$ was considered, wider with respect to that used for non-strange D mesons~\cite{ALICE:2021rxa} to cover possible yet unmeasured differences between the modification of charm- and beauty-baryon production in \PbPb collisions with respect to the one in \pp collisions.

The sources of systematic uncertainty considered in this analysis are assumed to be uncorrelated among each other and the total systematic uncertainty in each \pt and centrality interval is calculated as the quadratic sum of the individual uncertainties. For the \LcD ratio, the \Lc and \Dzero uncertainties were considered as uncorrelated except for the tracking efficiency and the feed-down contribution, which are assumed correlated and thus partially cancel in the ratio, and the systematic uncertainty of the centrality interval definition, which fully cancels. For the \RAA, the \pp and \PbPb uncertainties were considered as uncorrelated except for the branching ratio uncertainty and the feed-down contribution, which both partially cancel out (the former because the \pp measurement considers additional decay modes). Finally, in case of the \pt-integrated \LcD ratio, there is a correlation between the extrapolation uncertainty of the \Lc baryon and the measured uncertainties of the \Lc and \Dzero hadrons. To treat this correlation, the extrapolation uncertainty is divided into a correlated part (estimated as the extrapolation uncertainty when considering only the shape predicted by TAMU) and an uncorrelated part (the total extrapolation uncertainty subtracting the correlated part) with respect to the measured uncertainties. The uncorrelated part is summed in quadrature with the measured uncertainties, while the correlated part is added linearly.

%% file: 4_Results.tex
\section{Results}
\begin{figure*}[tb!]
\begin{center}
\includegraphics[width=0.48\textwidth]{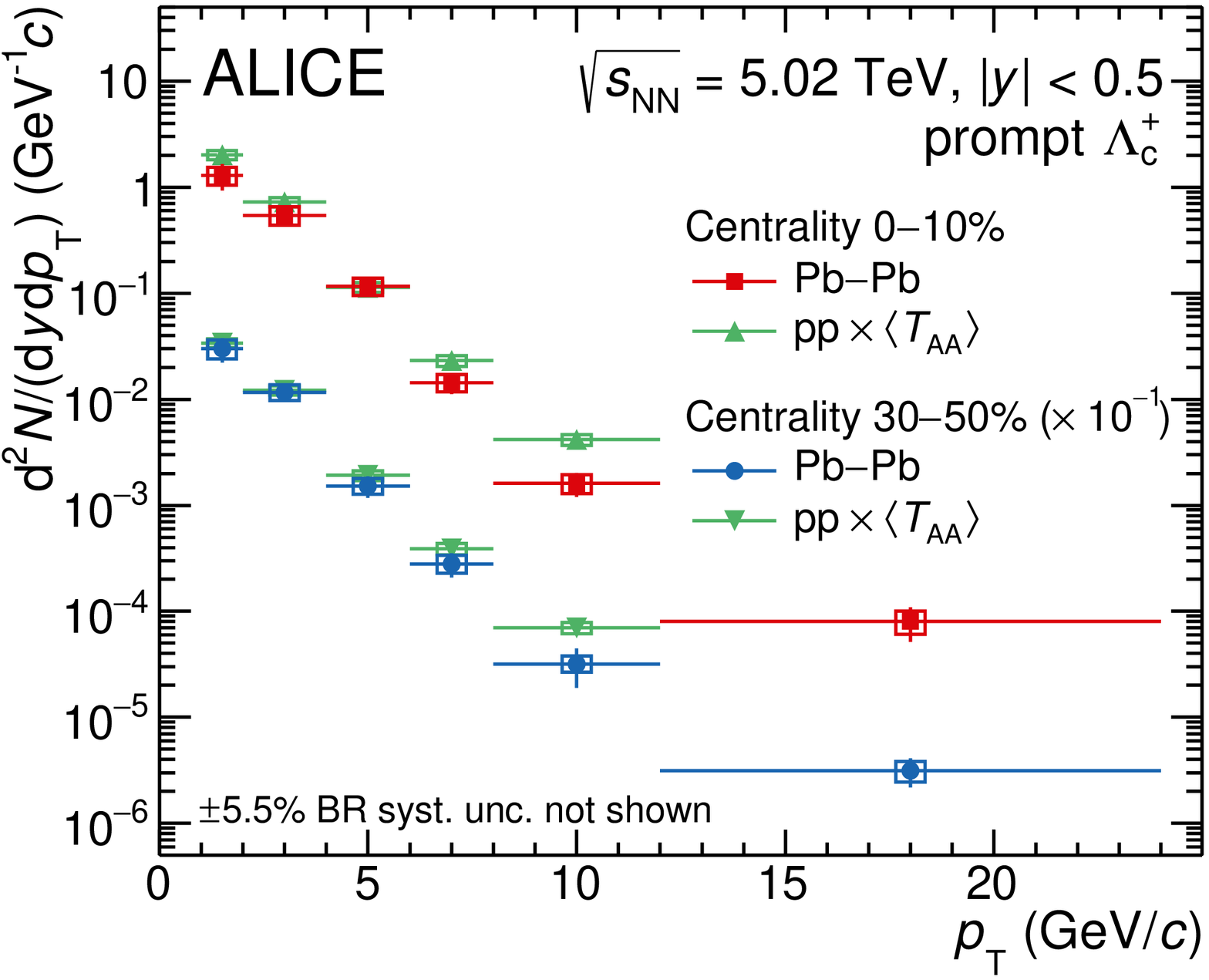}%
\includegraphics[width=0.48\textwidth]{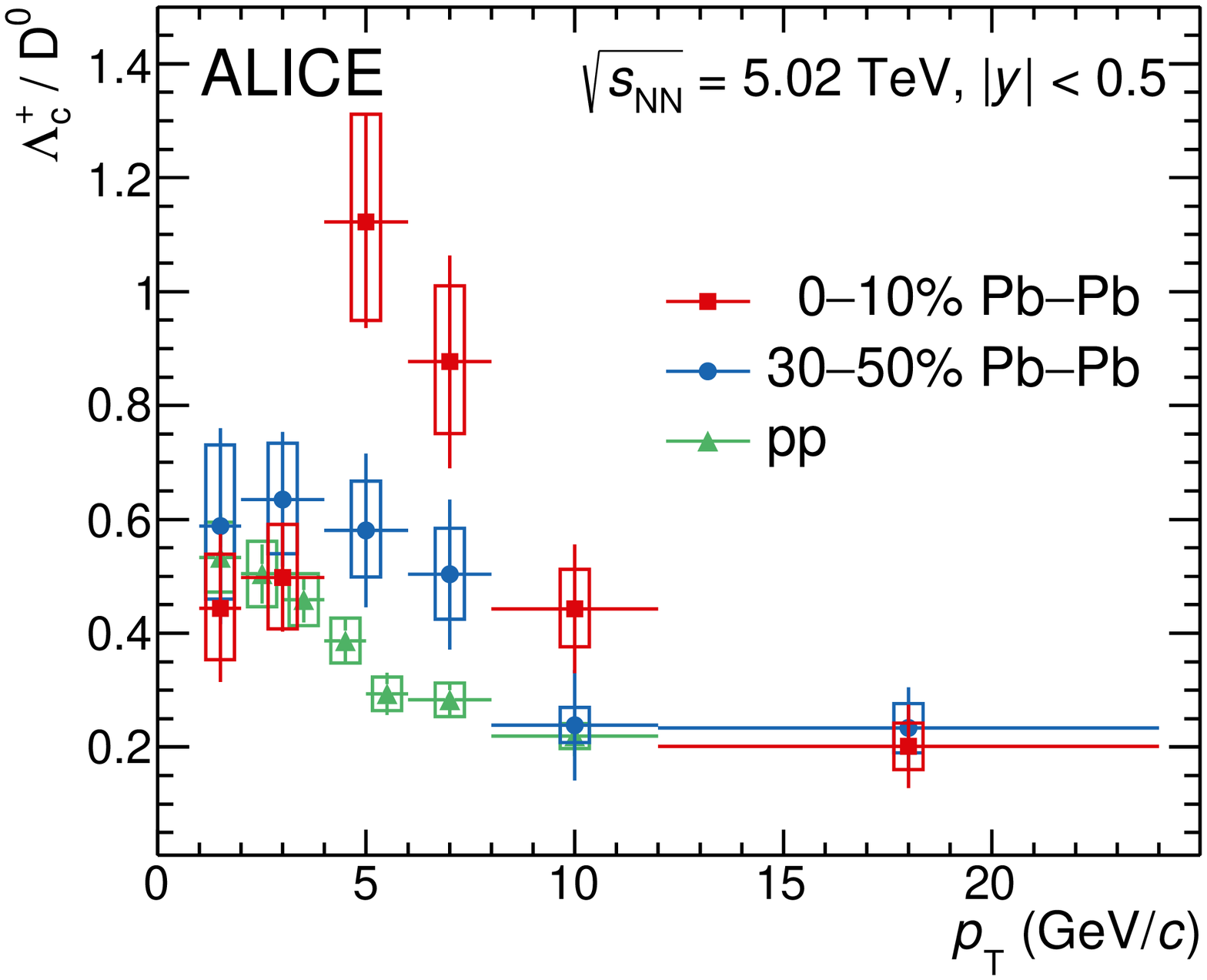}
\caption{Left: \pt-differential production yields of prompt \Lc in central (0--10\%) and mid-central (30--50\%) \PbPb collisions at \fivenn compared to the \pp reference~\cite{Acharya:2020lrg} scaled by the \TAA of the corresponding centrality interval~\cite{ALICE:2018tvk}. Right: \LcD ratio in central and mid-central \PbPb collisions at \fivenn compared with the results obtained from \pp collisions~\cite{Acharya:2020lrg}.}
\label{fig:spectra_ratios}
\end{center}
\end{figure*}

The \pt-differential production yields of prompt \Lc baryons are shown in Fig.~\ref{fig:spectra_ratios} (left panel). The statistical and total systematic uncertainties are shown as uncertainty bars and boxes, respectively, for all figures. The results are compared with the \pp reference cross section~\cite{Acharya:2020lrg}
multiplied by the corresponding \TAA~\cite{ALICE:2018tvk}, i.e.~the denominator of the \RAA observable that is discussed later. In the right panel of Fig.~\ref{fig:spectra_ratios}, the ratio of the production yields of \Lc baryons to that of \Dzero mesons, measured in the same centrality intervals~\cite{ALICE:2021rxa}, are presented together with the \pp measurement at the same collision energy~\cite{Acharya:2020lrg}. The ratios increase from \pp to mid-central and central \PbPb collisions for $4<\pt<8$~\gevc with a significance of $2.0$ and $3.7$ standard deviations, respectively. This trend is qualitatively similar to what is observed for the $\rm p/\pi$~\cite{Acharya:2019yoi} and $\Lambda/\kzero$~\cite{Abelev:2013xaa} ratios, which both show a distinct peak at intermediate \pt that increases in magnitude (by about a factor 2 for mid-central and a factor 3 for central \PbPb collisions with respect to minimum-bias \pp collisions) and shifts to higher \pt values (from about 2~\gevc in \pp to 4~\gevc in central \PbPb collisions) with increasing multiplicity. The central and mid-central \LcD ratios in $12<\pt<24$~\gevc are compatible with the measurement by CMS in 0--100\% \PbPb collisions in $\pt>10$~\gevc region~\cite{Sirunyan:2019fnc}. The central \LcD ratio in $6<\pt<8$~\gevc is in agreement with the previous measurement of ALICE in the 0--80\% centrality interval~\cite{Acharya:2018ckj}. For $\pt>4$~\gevc, the ratio measured in central collisions resembles in magnitude and \pt trend the one reported by STAR in $2.5<\pt<8$~\gevc in 10--80\% \AuAu collisions at \twoHnn~\cite{Adam:2019hpq}. Note that the large centrality classes of the previous measurements are dominated by the production in the most central events (given the scaling of the \Lc yields with $N_{\rm coll}\times\Raa$), hence they are compared to the measurement in 0--10\%.

\begin{figure*}[tb!]
  \begin{center}
    \includegraphics[trim={0.4cm 0 0.4cm 0},clip,width=0.96\textwidth]{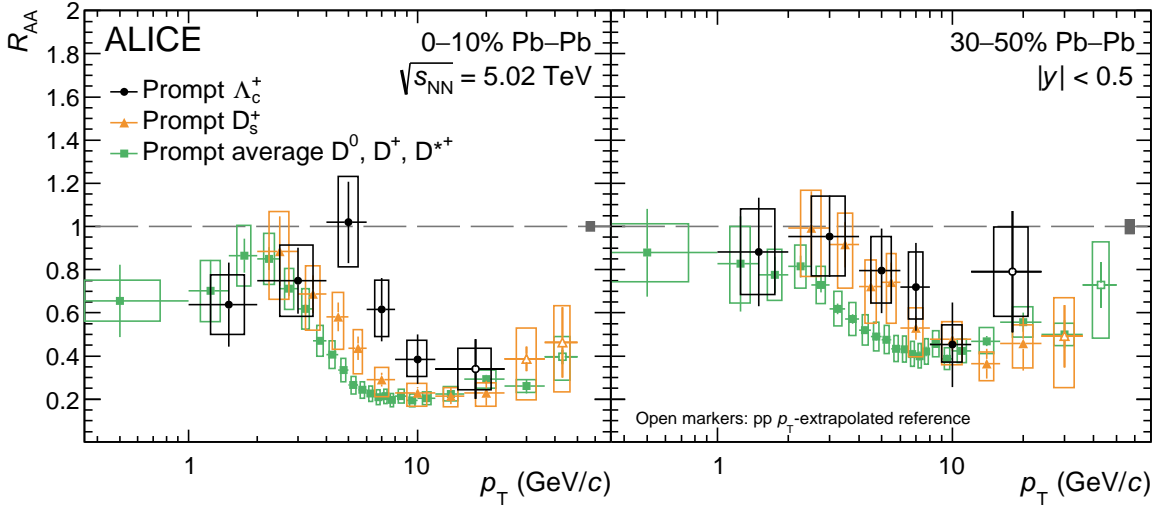}
    \caption{Nuclear modification factor \Raa of prompt \Lc baryons in central (0--10\%; left) and mid-central (30--50\%; right) \PbPb collisions at \fivenn, compared with the \Raa of prompt \Ds~\cite{ALICE:2021kfc} and the average of prompt non-strange $\rm D$ mesons~\cite{ALICE:2021rxa}. The normalization uncertainties are shown as boxes around unity.}
   \label{fig:RAAvsD}
  \end{center}
\end{figure*}

The nuclear modification factor \RAA of prompt \Lc is compared with the \Raa of prompt \Ds mesons~\cite{ALICE:2021kfc} and the average \Raa of prompt \Dzero, \Dplus, and \Dstar mesons~\cite{ALICE:2021rxa} in Fig.~\ref{fig:RAAvsD} for the 0--10\% and 30--50\% centrality intervals. The \pt-differential \Lc cross section in \pp collisions at \five in the $1<\pt<12$~\gevc interval from~\cite{Acharya:2020lrg} was used as the \pp reference. In the interval $12<\pt<24$~\gevc, the \Lc and \Dzero measurements at $\sqrt{s}=5.02$ and 13~TeV~\cite{Acharya:2019mgn,Acharya:2021vpo} were exploited, assuming no \sqrts dependence for the \LcD ratio as observed within uncertainties in $1<\pt<12$~\gevc~\cite{Acharya:2021vpo}. The total uncertainty of the \pp reference in the $12<\pt<24$~\gevc interval is 23\%, combining in quadrature the measured statistical and systematic uncertainties on the \LcD ratio at \thirteen and \Dzero cross section at \five. 

The suppression of all charm-meson (baryon) species from $\pt\gtrsim3\,(6)$~\gevc is understood as being primarily due to the interaction of charm quarks with the quark--gluon plasma, which modifies their momentum spectra, as discussed extensively for the non-strange $\rm D$ mesons in~\cite{ALICE:2021rxa}. In central collisions in the region $4 < \pt < 8$~\gevc, there is a hint of a hierarchy $\Raa(\rm D) < \Raa(\Ds) < \Raa(\Lc)$. In mid-central collisions, this hierarchy is less pronounced. In the $\pt \gtrsim 10$~\gevc region, where the hadronization is expected to occur mainly via fragmentation, the \RAA of the various charm-hadron species are compatible within uncertainties.

\begin{figure*}[tb!]
\begin{center}
\includegraphics[width=\textwidth]{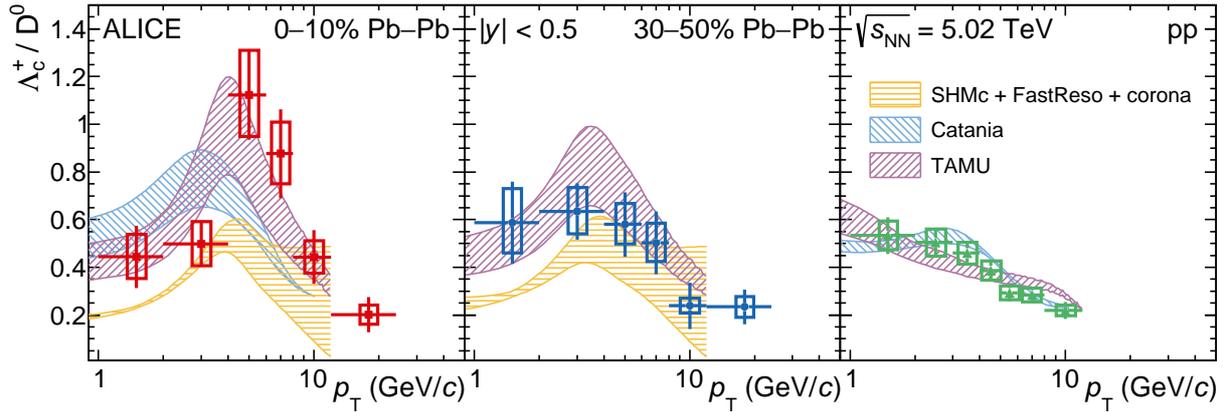}
\caption{The \LcD yield ratio as a function of \pt in 0--10\% (left) and 30--50\% (middle) \PbPb and \pp (right) collisions at \fivenn compared with predictions of different theoretical calculations~\cite{Plumari:2017ntm,He:2019vgs,Andronic:2021erx,Mazeliauskas:2018irt,Minissale:2020bif,He:2019tik}.}
\label{fig:LcD0theory}
\end{center}
\end{figure*}

Figure~\ref{fig:LcD0theory} compares the \pt-differential \LcDzero ratios with different theoretical predictions: Catania~\cite{Plumari:2017ntm}, TAMU~\cite{He:2019vgs}, and the GSI--Heidelberg statistical hadronization model (SHMc)~\cite{Andronic:2021erx}. The predictions of Catania and TAMU for \pp collisions~\cite{Minissale:2020bif,He:2019tik} are also compared with the existing measurement in \pp collisions~\cite{Acharya:2020lrg}. The Catania model~\cite{Plumari:2017ntm,Minissale:2020bif} assumes that a QGP is formed in both \pp and \PbPb collisions. In \PbPb collisions heavy-quark transport is implemented via the Boltzmann equation, and in both \pp and \PbPb collisions hadronization occurs either via coalescence, implemented through the Wigner formalism, or via fragmentation in case the quarks do not undergo coalescence. The TAMU model~\cite{He:2019vgs} describes charm-quark transport in an expanding medium with the Langevin equation and hadronization proceeds primarily via coalescence, implemented with a Resonance Recombination Model (RRM)~\cite{Ravagli:2007xx}. Left-over charm quarks not undergoing coalescence are hadronized via fragmentation. In \pp collisions, the charm-hadron abundances are instead determined with a statistical hadronization approach~\cite{He:2019tik}. In both collision systems the underlying charm-baryon spectrum includes unobserved excited states~\cite{Zyla:2020zbs} predicted by the Relativistic Quark Model (RQM)~\cite{Ebert:2011kk} and lattice QCD~\cite{He:2019tik}. Finally, for the SHMc predictions~\cite{Andronic:2021erx}, which include only charm mesons and baryons established experimentally, the charm-hadron \pt spectra are modeled within a core-corona approach. The core contribution represents the central region of the colliding nuclei where charm quarks achieve local thermal equilibrium in a hydrodynamically expanding QGP. The charm-hadron spectra in the corona contribution are, instead, parameterized from measurements in \pp collisions. The \pt-spectra modification due to resonance decays is computed using the FastReso package~\cite{Mazeliauskas:2018irt}. The theoretical uncertainty bands shown in Fig.~\ref{fig:LcD0theory} derive from: an assumed range of branching ratios (50--100\%) for the decays of the RQM-augmented excited states into \Lc for the TAMU model; the variation of about 10\% of the Wigner function widths in the Catania calculations; and mainly the uncertainties on the \pp spectra fits in the SHMc predictions at high \pt.

The SHMc describes the \LcD ratio in mid-central collisions, but underpredicts the ratio in $4 < \pt < 8$~\gevc in central collisions by about $2.5\sigma$ of the combined statistical, systematic, and theoretical uncertainties. The prediction of the Catania model in central collisions underestimates the \LcD ratio at intermediate \pt, although the deviation is at maximum $2.5\sigma$. The TAMU predictions reproduce the magnitude and shape of the \LcD ratios. While both these fragmentation plus coalescence model calculations are able to describe the \LcD ratio in \AuAu collisions at \twoHnn in the 10--80\% centrality interval~\cite{Adam:2019hpq}, the TAMU model better reproduces the data in central \PbPb collisions. A pure coalescence scenario from an older version of the Catania model was reproducing better the previous ALICE measurement in 0--80\% \PbPb collisions~\cite{Acharya:2018ckj}. The Catania and TAMU predictions also describe both the magnitude and \pt shape of the measured \LcD ratio in \pp collisions. Instead, at forward rapidity, the TAMU model predicts a systematically higher \LcD ratio than measured by LHCb in 65--90\% \PbPb collisions at \fivenn~\cite{LHCb:2022ddg}.

The \Lc production yield for $\pt>0$ was estimated by summing up the measured \pt-differential yields and the extrapolated \Lc yield for $\pt<1$~\gevc. The \Lc yield in $0<\pt<1$~\gevc was obtained as the product of the \LcD ratio value estimated by interpolating the ratio in the measured \pt interval with model expectations and the measured \Dzero yield~\cite{ALICE:2021rxa}. The interpolation procedure was performed using the shape predicted by TAMU~\cite{He:2019vgs}, Catania~\cite{Plumari:2017ntm} (not available for 30--50\%), SHMc~\cite{Andronic:2021erx}, and blast-wave~\cite{Schnedermann:1993ws} calculations, leaving the normalization as a free parameter. The shape from TAMU was chosen as the central value based on the $\chi^2/{\rm ndf}$ values, while the difference between the obtained yields was considered in the systematic uncertainty due to the extrapolation. The results for the prompt \Lc production yields per unit of rapidity in $|y| < 0.5$ are ${\rm d}N/{\rm d}y = 3.27 \pm 0.42\,({\rm stat}) \pm 0.45\,({\rm syst}) \pm 0.16\,({\rm BR})\,^{+0.46}_{-0.29}\,({\rm extr})$ for central collisions and ${\rm d}N/{\rm d}y = 0.70 \pm 0.09\,({\rm stat}) \pm 0.09\,({\rm syst}) \pm 0.04\,({\rm BR})\,^{+0.07}_{-0.05}\,({\rm extr})$ for mid-central collisions, where the visible yield is about 81\% of the total for both centrality classes. The SHMc~\cite{Andronic:2021erx} predicts lower values, ${\rm d}N/{\rm d}y = 1.55 \pm 0.23$ and ${\rm d}N/{\rm d}y = 0.316 \pm 0.036$, respectively. 

\begin{figure}[tb!]
\begin{center}
\includegraphics[width=0.48\textwidth]{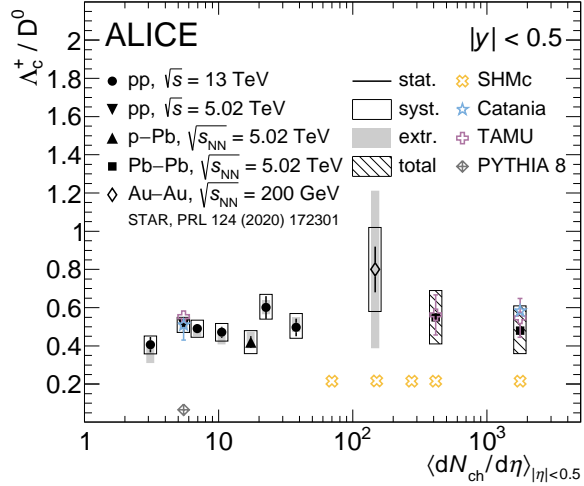}%
\caption{The \pt-integrated and to $\pt>0$ extrapolated \LcD ratios in central and mid-central \PbPb collisions at \fivenn compared to the same ratio at \pp and \pPb~\cite{Acharya:2020lrg,ALICE:2021npz} and \AuAu~\cite{Adam:2019hpq} multiplicities. Predictions from theoretical calculations are shown as well~\cite{He:2019vgs,Andronic:2021erx,He:2019tik,Sjostrand:2014zea,Minissale:2020bif,Plumari:2017ntm}.}
\label{fig:LcD0_ptint}
\end{center}
\end{figure}

The measured \LcD ratios, obtained dividing the \pt-integrated \Lc and \Dzero yields~\cite{ALICE:2021rxa}, are presented in Fig.~\ref{fig:LcD0_ptint}, taking into account the correlation between the measured and extrapolated uncertainties. Similarly to what is observed for the $\Lambda/\kzero$ ratio~\cite{ALICE:2019avo,Abelev:2013xaa}, the \LcD ratios in \PbPb collisions are compatible with the \pt-integrated \LcD ratios at \pp and \pPb multiplicities~\cite{Acharya:2020lrg,ALICE:2021npz} within one standard deviation of the combined uncertainties. This observation, together with the significant enhancement of the \LcD ratio at intermediate \pt with increasing multiplicity, seen here and in \pp collisions~\cite{ALICE:2021npz}, suggests a modified (and perhaps similar) mechanism of hadronization in all hadronic collision systems with respect to charm fragmentation tuned on \ee and \ep measurements (PYTHIA~8 point in Fig.~\ref{fig:LcD0_ptint}). The coalescence models of~\cite{Lee:2007wr,Oh:2009zj,Beraudo:2022dpz}, in which the \LcD ratio depends on the balance of quark and diquark densities at hadronization time, expect a dependence of the \pt-integrated \LcD ratio on multiplicity (leading to an increase by about a factor 3--10 in nuclear collisions compared with their \pp baseline), which is not observed. The measured \pt-differential enhancement may, instead, predominantly be caused by altered production ratios for baryons and mesons following from the phase-space distribution of the quarks. This can arise from the collective radial expansion of the system, for which, in the coalescence picture (Catania and TAMU \PbPb points in Fig.~\ref{fig:LcD0_ptint}), the accounting of space--momentum correlations in the procedure have been observed to be fundamental in~\cite{He:2019vgs,Beraudo:2022dpz}. Interactions in the hadronic phase are, on the contrary, expected to have a small effect on the \LcD ratio~\cite{Ghosh:2014oia,Das:2016llg}. The statistical hadronization approach (SHMc and TAMU \pp points in Fig.~\ref{fig:LcD0_ptint}), can also describe both the \pt-differential and \pt-integrated observations with the, currently debated, caveat that for the proper normalization yet unobserved charm-baryon states need to be assumed~\cite{He:2019tik,Andronic:2021erx}. Note that the authors of the TAMU model include these additional states already in their predictions, while for the SHMc model it is not the baseline. The uncertainty of the \pt-integrated yield in \PbPb collisions is still relatively large, and more precise measurements at low \pt will help to further discriminate between charm-baryon formation scenarios.

\begin{figure*}[tb!]
  \begin{center}
    \includegraphics[trim={0.4cm 0 0.4cm 0},clip,width=0.96\textwidth]{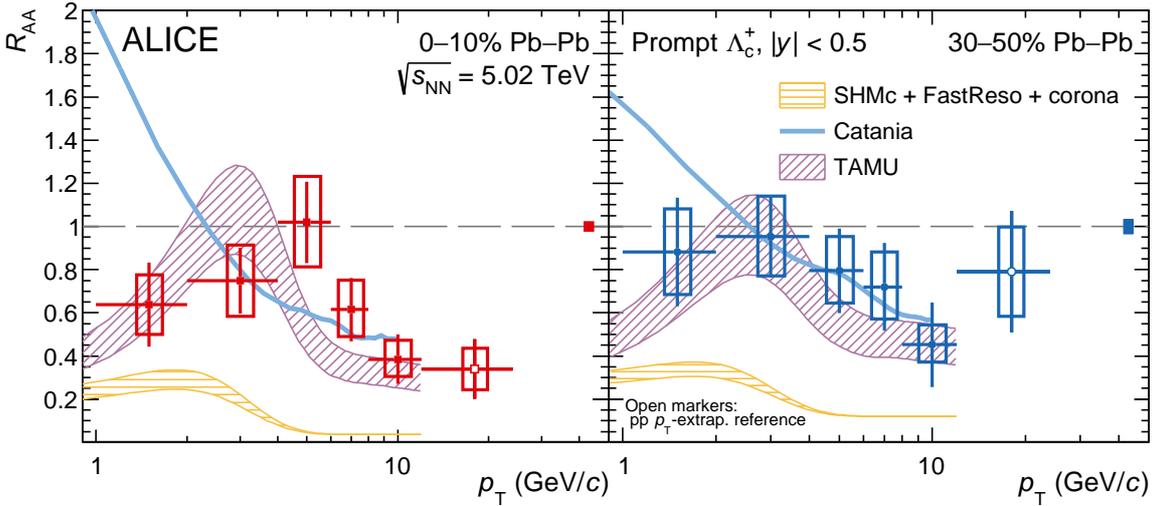}
    \caption{Nuclear modification factor \Raa of prompt \Lc baryons in central (0--10\%; left) and mid-central (30--50\%; right) \PbPb collisions at \fivenn, compared with model predictions. When estimated, the model uncertainty is shown as a shaded band.}
   \label{fig:RAAvsTheory}
  \end{center}
\end{figure*}

Finally, Fig.~\ref{fig:RAAvsTheory} shows the \Raa of prompt \Lc baryons compared with the previously introduced theoretical models~\cite{Plumari:2017ntm,He:2019vgs,Andronic:2021erx}. The Catania \RAA predictions are from an earlier version of the model than the \LcD predictions and they do not have an uncertainty band. The TAMU model provides a good description of the \Raa, over the whole \pt range, in both central and mid-central collisions. The Catania model describes the data in both central and mid-central collisions for $\pt>2$~\gevc, however for $\pt<2$~\gevc the model predicts a \RAA higher than unity which is disfavored by data. Both these models do not include charm-quark interactions with medium constituents via radiative processes, hence are not expected to describe the \Raa for $\pt > 8$~\gevc. The SHMc model instead significantly underestimates the \Lc \Raa over the whole \pt range.

%% file: 5_Conclusion.tex
\section{Conclusions}

In summary, the measurements of the production yield of prompt \Lambdac baryons in central (0--10\%) and mid-central (30--50\%) \PbPb collisions at a center-of-mass energy per nucleon pair \fivenn were presented. The yield could be extrapolated to $\pt=0$ in the two centrality classes with significantly smaller uncertainties than the previous measurement by STAR in 10--80\% \AuAu collisions at \twoHnn, exploring not only a new energy regime but also higher multiplicities. The \pt-differential \LcD ratios increase from \pp to central \PbPb collisions for $4 < \pt < 8$~\gevc with a significance of $3.7$ standard deviations, while the \pt-integrated ratios are compatible within one standard deviation. Both observations are in qualitative agreement with the baryon-to-meson ratio for strange hadrons. The measurements are described by theoretical calculations that include both coalescence and fragmentation processes when describing the hadronization of heavy flavors in the QGP. The upgraded ALICE detector for the LHC Runs 3 and 4 will increase its acquisition rate by up to a factor of about 50 in \PbPb collisions and the tracking precision by a factor 3--6, meaning future measurements of \Lc-baryon production will allow for stronger constraints on the heavy-quark hadronization mechanisms in heavy-ion collisions~\cite{Citron:2018lsq}.

%% file: fa_2021-11-30.tex
% Version: 2021-11-30

The ALICE Collaboration would like to thank all its engineers and technicians for their invaluable contributions to the construction of the experiment and the CERN accelerator teams for the outstanding performance of the LHC complex.
The ALICE Collaboration gratefully acknowledges the resources and support provided by all Grid centres and the Worldwide LHC Computing Grid (WLCG) collaboration.
The ALICE Collaboration acknowledges the following funding agencies for their support in building and running the ALICE detector:
A. I. Alikhanyan National Science Laboratory (Yerevan Physics Institute) Foundation (ANSL), State Committee of Science and World Federation of Scientists (WFS), Armenia;
Austrian Academy of Sciences, Austrian Science Fund (FWF): [M 2467-N36] and Nationalstiftung f\"{u}r Forschung, Technologie und Entwicklung, Austria;
Ministry of Communications and High Technologies, National Nuclear Research Center, Azerbaijan;
Conselho Nacional de Desenvolvimento Cient\'{\i}fico e Tecnol\'{o}gico (CNPq), Financiadora de Estudos e Projetos (Finep), Funda\c{c}\~{a}o de Amparo \`{a} Pesquisa do Estado de S\~{a}o Paulo (FAPESP) and Universidade Federal do Rio Grande do Sul (UFRGS), Brazil;
Ministry of Education of China (MOEC) , Ministry of Science \& Technology of China (MSTC) and National Natural Science Foundation of China (NSFC), China;
Ministry of Science and Education and Croatian Science Foundation, Croatia;
Centro de Aplicaciones Tecnol\'{o}gicas y Desarrollo Nuclear (CEADEN), Cubaenerg\'{\i}a, Cuba;
Ministry of Education, Youth and Sports of the Czech Republic, Czech Republic;
The Danish Council for Independent Research | Natural Sciences, the VILLUM FONDEN and Danish National Research Foundation (DNRF), Denmark;
Helsinki Institute of Physics (HIP), Finland;
Commissariat \`{a} l'Energie Atomique (CEA) and Institut National de Physique Nucl\'{e}aire et de Physique des Particules (IN2P3) and Centre National de la Recherche Scientifique (CNRS), France;
Bundesministerium f\"{u}r Bildung und Forschung (BMBF) and GSI Helmholtzzentrum f\"{u}r Schwerionenforschung GmbH, Germany;
General Secretariat for Research and Technology, Ministry of Education, Research and Religions, Greece;
National Research, Development and Innovation Office, Hungary;
Department of Atomic Energy Government of India (DAE), Department of Science and Technology, Government of India (DST), University Grants Commission, Government of India (UGC) and Council of Scientific and Industrial Research (CSIR), India;
Indonesian Institute of Science, Indonesia;
Istituto Nazionale di Fisica Nucleare (INFN), Italy;
Japanese Ministry of Education, Culture, Sports, Science and Technology (MEXT) and Japan Society for the Promotion of Science (JSPS) KAKENHI, Japan;
Consejo Nacional de Ciencia (CONACYT) y Tecnolog\'{i}a, through Fondo de Cooperaci\'{o}n Internacional en Ciencia y Tecnolog\'{i}a (FONCICYT) and Direcci\'{o}n General de Asuntos del Personal Academico (DGAPA), Mexico;
Nederlandse Organisatie voor Wetenschappelijk Onderzoek (NWO), Netherlands;
The Research Council of Norway, Norway;
Commission on Science and Technology for Sustainable Development in the South (COMSATS), Pakistan;
Pontificia Universidad Cat\'{o}lica del Per\'{u}, Peru;
Ministry of Education and Science, National Science Centre and WUT ID-UB, Poland;
Korea Institute of Science and Technology Information and National Research Foundation of Korea (NRF), Republic of Korea;
Ministry of Education and Scientific Research, Institute of Atomic Physics, Ministry of Research and Innovation and Institute of Atomic Physics and University Politehnica of Bucharest, Romania;
Joint Institute for Nuclear Research (JINR), Ministry of Education and Science of the Russian Federation, National Research Centre Kurchatov Institute, Russian Science Foundation and Russian Foundation for Basic Research, Russia;
Ministry of Education, Science, Research and Sport of the Slovak Republic, Slovakia;
National Research Foundation of South Africa, South Africa;
Swedish Research Council (VR) and Knut \& Alice Wallenberg Foundation (KAW), Sweden;
European Organization for Nuclear Research, Switzerland;
Suranaree University of Technology (SUT), National Science and Technology Development Agency (NSDTA) and Office of the Higher Education Commission under NRU project of Thailand, Thailand;
Turkish Energy, Nuclear and Mineral Research Agency (TENMAK), Turkey;
National Academy of  Sciences of Ukraine, Ukraine;
Science and Technology Facilities Council (STFC), United Kingdom;
National Science Foundation of the United States of America (NSF) and United States Department of Energy, Office of Nuclear Physics (DOE NP), United States of America.

%% file: 6_SupplementalFigures.tex
\section{Raw-yield extraction}
\label{app:Mass}
\begin{figure*}[h!]
  \begin{center}
    \includegraphics[width=0.47\textwidth]{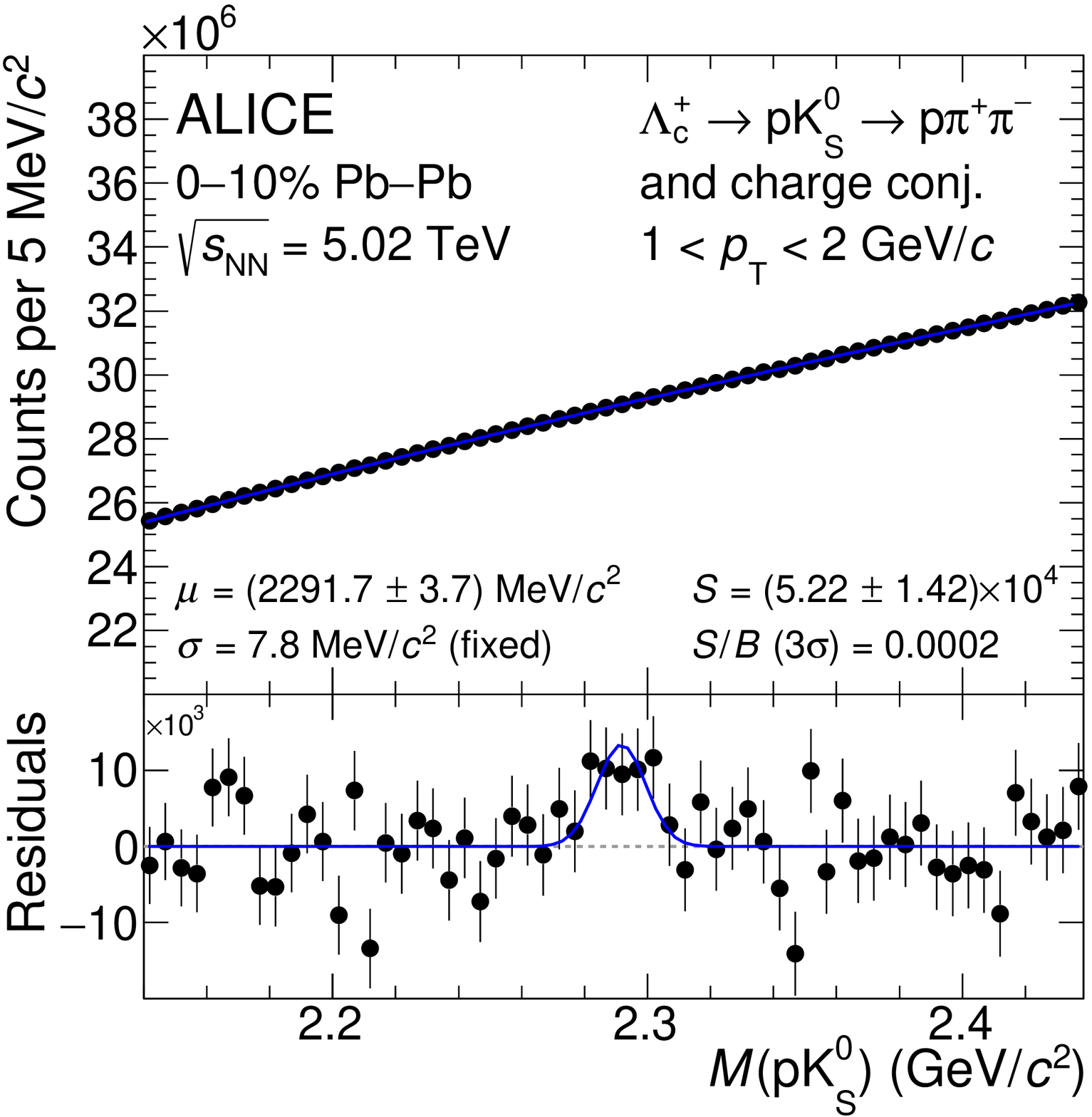}%
    \hspace{0.3cm}
    \includegraphics[width=0.47\textwidth]{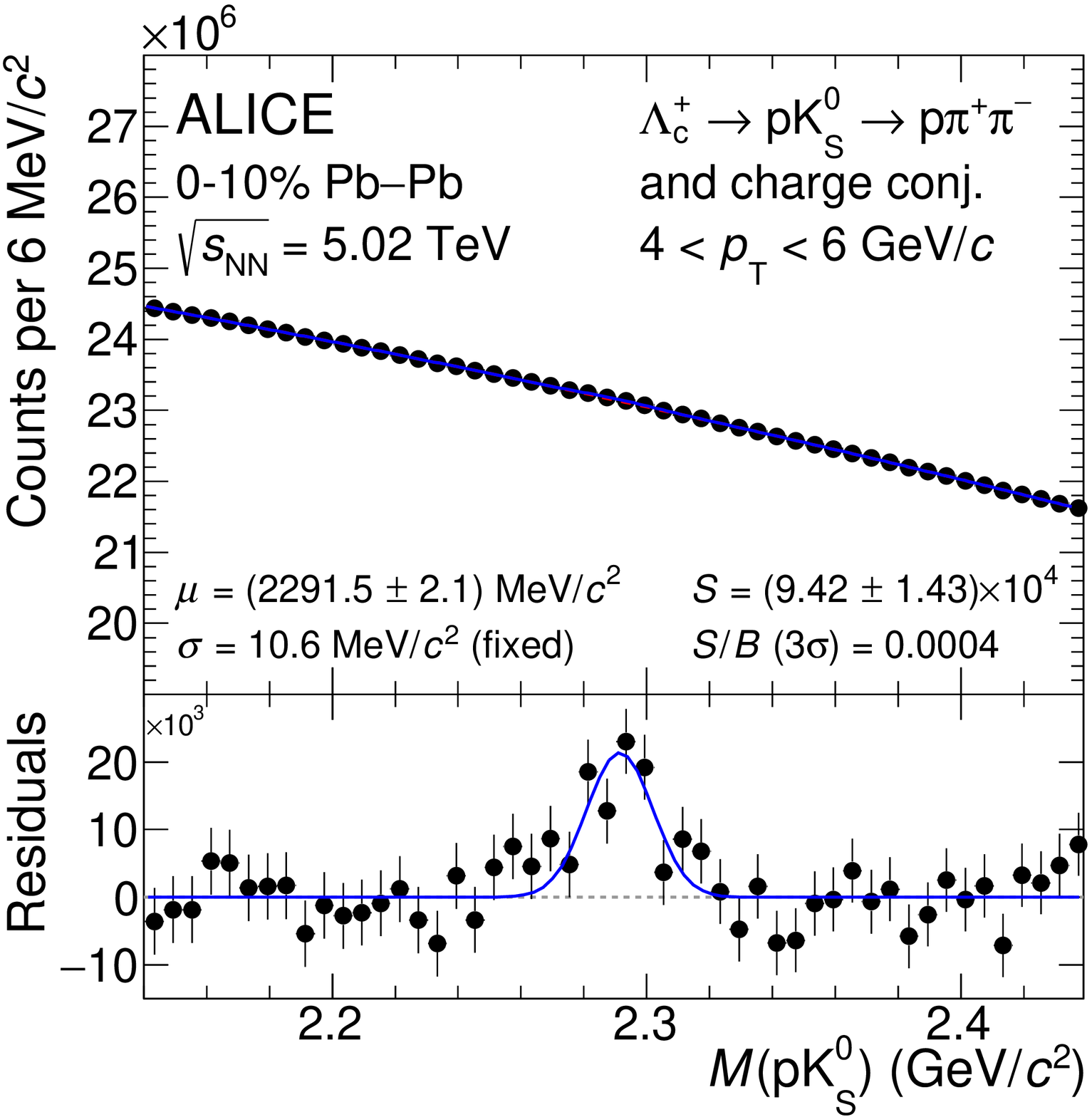}
    \includegraphics[width=0.47\textwidth]{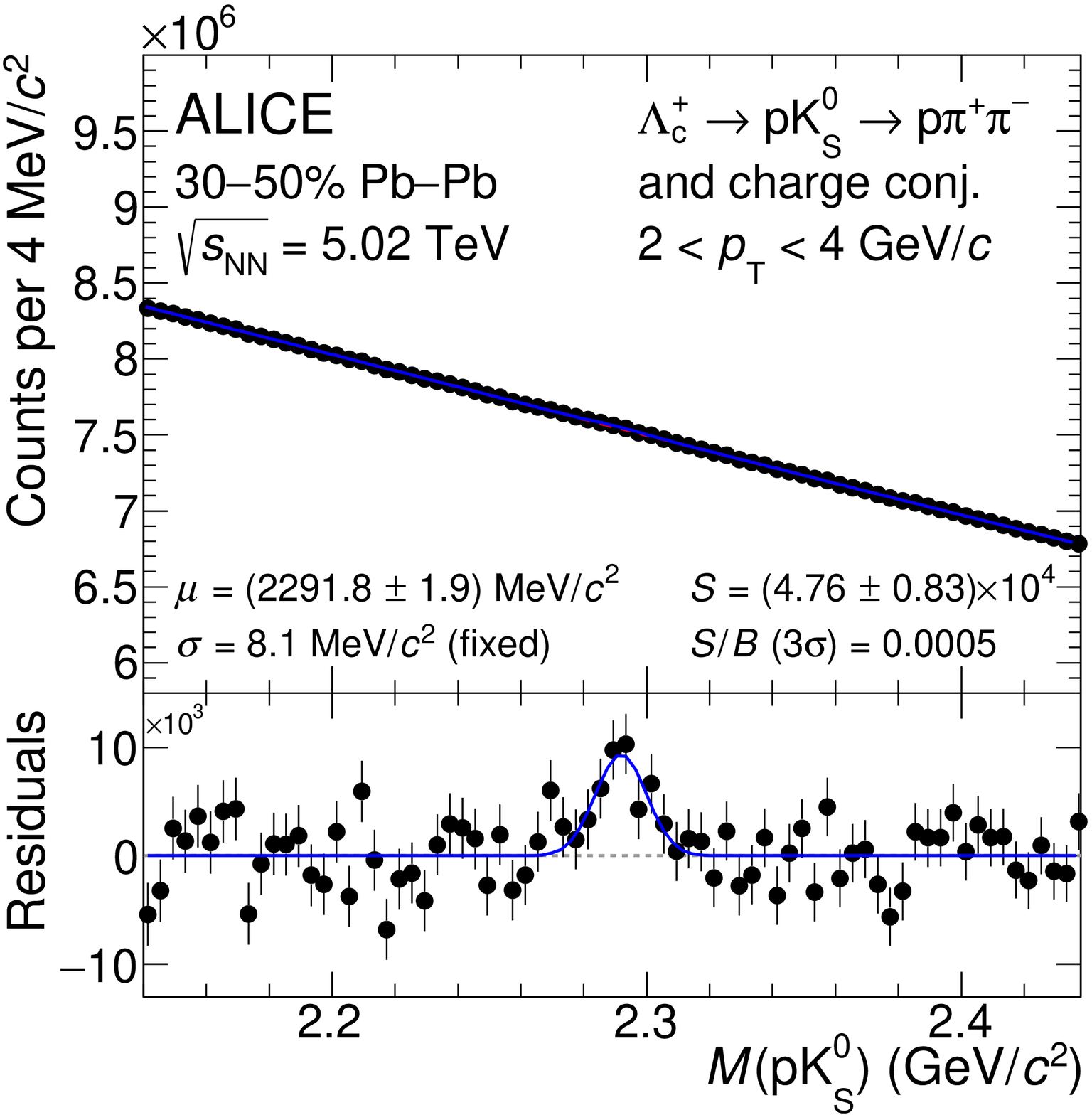}%
    \hspace{0.3cm}
    \includegraphics[width=0.47\textwidth]{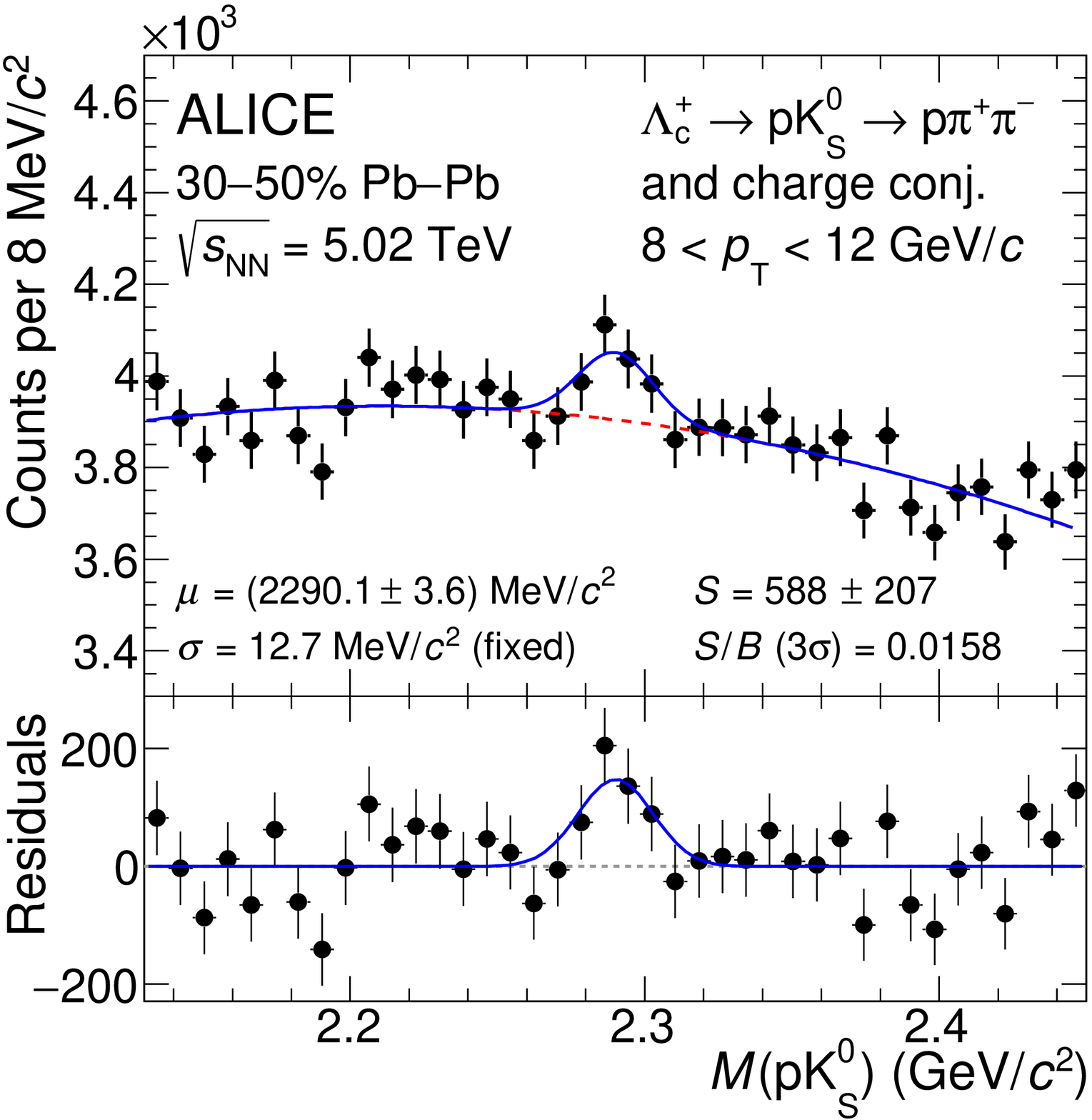}
    \caption{Invariant mass ($M$) distributions of \LctopKzerosfull candidates and charge conjugates in different \pt intervals in central (0--10\%; top) and mid-central (30--50\%; bottom) \PbPb collisions at \fivenn. The blue solid lines show the total fit functions and the red dashed lines are the combinatorial-background terms. The values of the mean ($\mu$) and the width ($\sigma$) of the signal peak are reported together with the signal counts ($S$) and the signal-over-background ratio ($S/B$) in the mass interval ($\mu-3\sigma,\mu+3\sigma$).}
   \label{fig:InvMass}
  \end{center}
\end{figure*}

Examples of the invariant mass distributions from which the \Lc raw yields are extracted are reported in Fig.~\ref{fig:InvMass}. The spectra together with the result of the fits in $1<\pt<2$~\gevc and $4<\pt<6$~\gevc for central (0--10\%) and $2<\pt<4$~\gevc and $8<\pt<12$~\gevc for mid-central (30--50\%) \PbPb collisions are shown.

%% file: 2021-11-30-Alice_Authorlist_2021-11-30.tex
\begin{flushleft}

\bigskip 

S.~Acharya$^{\rm 142}$, 
D.~Adamov\'{a}$^{\rm 96}$, 
A.~Adler$^{\rm 74}$, 
J.~Adolfsson$^{\rm 81}$, 
G.~Aglieri Rinella$^{\rm 34}$, 
M.~Agnello$^{\rm 30}$, 
N.~Agrawal$^{\rm 54}$, 
Z.~Ahammed$^{\rm 142}$, 
S.~Ahmad$^{\rm 16}$, 
S.U.~Ahn$^{\rm 76}$
% ...

\bigskip

\bigskip 

\textbf{\Large Affiliation Notes}

\bigskip 

% ...
$^{\rm VII}$ Also at: University of Kansas, Lawrence, Kansas, United States\\

\bigskip

\bigskip 

\textbf{\Large Collaboration Institutes}

\bigskip 

% ...
$^{16}$ Department of Physics, Aligarh Muslim University, Aligarh, India\\
% ...
$^{30}$ Dipartimento DISAT del Politecnico and Sezione INFN, Turin, Italy\\
% ...
$^{34}$ European Organization for Nuclear Research (CERN), Geneva, Switzerland\\
% ...
$^{54}$ INFN, Sezione di Bologna, Bologna, Italy\\
% ...
$^{74}$ Johann-Wolfgang-Goethe Universit\"{a}t Frankfurt Institut f\"{u}r Informatik, Fachbereich Informatik und Mathematik, Frankfurt, Germany\\
% ...
$^{76}$ Korea Institute of Science and Technology Information, Daejeon, Republic of Korea\\
% ...
$^{81}$ Lund University Department of Physics, Division of Particle Physics, Lund, Sweden\\
% ...
$^{96}$ Nuclear Physics Institute of the Czech Academy of Sciences, \v{R}e\v{z} u Prahy, Czech Republic\\
% ...
$^{142}$ Variable Energy Cyclotron Centre, Homi Bhabha National Institute, Kolkata, India\\
% ...

\bigskip 

\end{flushleft}